\documentclass[showpacs,amsmath,amssymb,aps,showkeys,floatfix,prd,a4paper]{revtex4}

\usepackage[dvips]{graphicx}
\usepackage{dcolumn}
\usepackage{bm}
\usepackage{epsfig}
\usepackage{amsfonts}
\usepackage{amssymb,amscd}

\def\lsim{\raise0.3ex\hbox{$<$\kern-0.75em\raise-1.1ex\hbox{$\sim$}}}

\def\gsim{\raise0.3ex\hbox{$>$\kern-0.75em\raise-1.1ex\hbox{$\sim$}}}

\newcommand{\be}{\begin{equation}}

\newcommand{\ee}{\end{equation}}

\def\beq{\begin{equation}}

\def\eeq{\end{equation}}

\def\beqa{\begin{eqnarray}}

\def\eeqa{\end{eqnarray}}

\newcommand{\rd}{\mbox{\boldmath $\Delta$}}

\newcommand{\ba}{\begin{eqnarray}}

\newcommand{\rr}{\mbox{\boldmath $r$}}

\newcommand{\rb}{\mbox{\boldmath $b$}}

\def\gappeq{\mathrel{\rlap {\raise.5ex\hbox{$>$}}

{\lower.5ex\hbox{$\sim$}}}}

\def\lappeq{\mathrel{\rlap{\raise.5ex\hbox{$<$}}

{\lower.5ex\hbox{$\sim$}}}}

\def\Toprel#1\over#2{\mathrel{\mathop{#2}\limits^{#1}}}

\begin{document}

\begin{flushright}
LU TP 16-XX\\
April 2016
\vskip1cm
\end{flushright}

\title{Double vector meson production in photon - hadron interactions at hadronic colliders}
\author{V.P. Gon\c{c}alves $^{1,2}$,  B.D.  Moreira$^{3}$  and  F.S. Navarra$^3$}
\affiliation{$^1$ Department of Astronomy and Theoretical Physics, Lund University, SE-223 62 Lund, Sweden \\  
$^{2}$ High and Medium Energy Group, Instituto de F\'{\i}sica e Matem\'atica,  Universidade Federal de Pelotas\\
Caixa Postal 354,  96010-900, Pelotas, RS, Brazil.\\
$^3$Instituto de F\'{\i}sica, Universidade de S\~{a}o Paulo,
C.P. 66318,  05315-970 S\~{a}o Paulo, SP, Brazil\\
}

\begin{abstract}
In this paper we analyse the double vector meson production in  photon -- hadron ($\gamma h$) interactions at $pp/pA/AA$ collisions and present predictions for the $\rho\rho$, $J/\Psi J/\Psi$ and $\rho J/\Psi$ production considering the double scattering mechanism. We estimate the total cross sections and rapidity distributions at LHC energies and compare our results with the predictions for the double vector meson production in $\gamma \gamma$ interactions at hadronic colliders.
We present predictions for the different rapidity ranges probed by the ALICE, ATLAS, CMS and LHCb Collaborations. Our results demonstrate that the $\rho\rho$ and  $J/\Psi J/\Psi$ production in $PbPb$ collisions is dominated by the double scattering mechanism, while the two - photon mechanism dominates in $pp$ collisions. Moreover, our results indicate that the analysis of the $\rho J/\Psi$ production at LHC can be useful to constrain the double scattering mechanism.
\end{abstract}



\maketitle

\vspace{1cm}

\begin{figure}
%\centerline{\psfig{figure=gamavec.eps,width=10cm}}  
{\psfig{figure=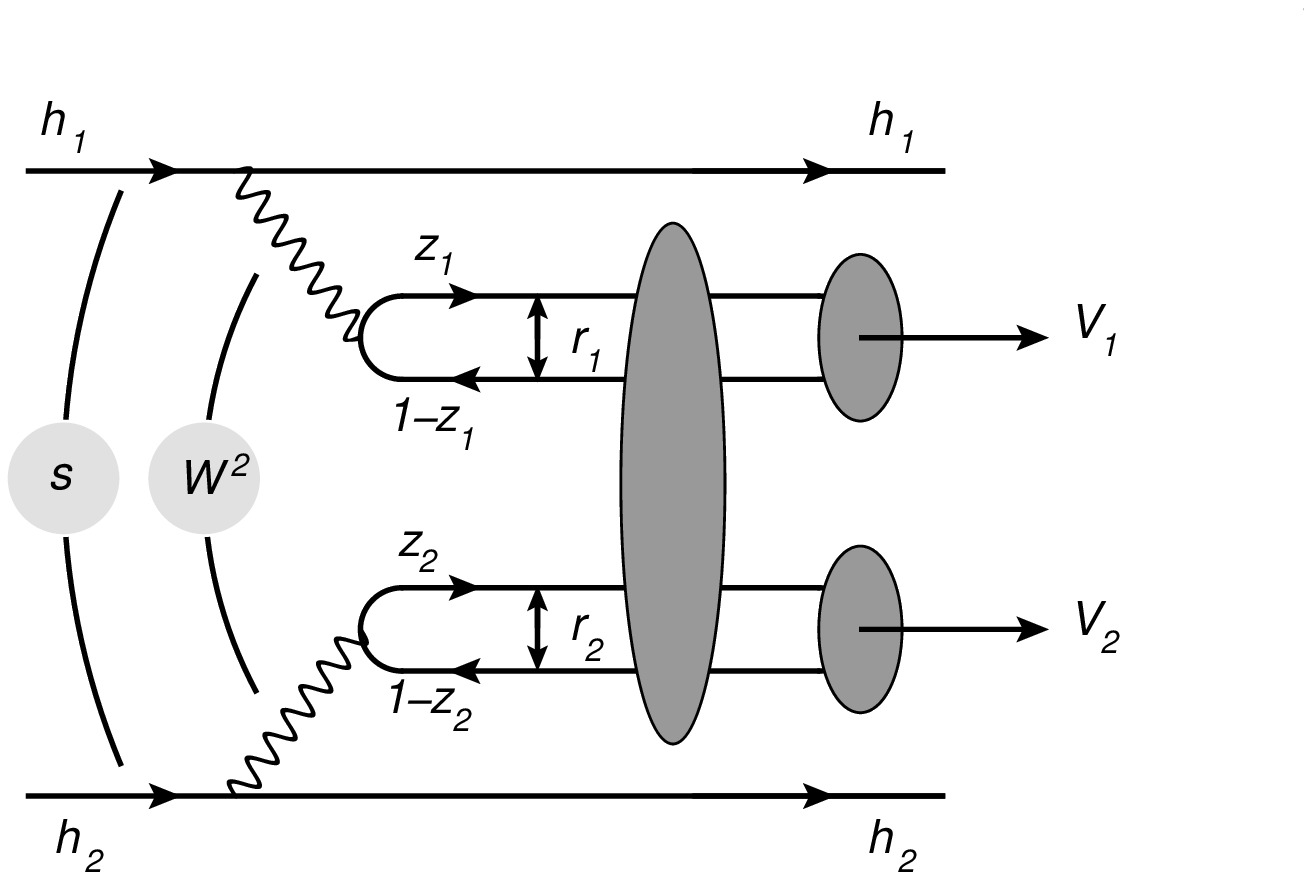,width=9cm}}
\caption{Double vector meson production in  $\gamma \gamma$ interactions at hadronic colliders  in the 
color dipole picture.}
\label{dia1}
\end{figure}

\begin{figure}
\begin{tabular}{ccc}
{\psfig{figure=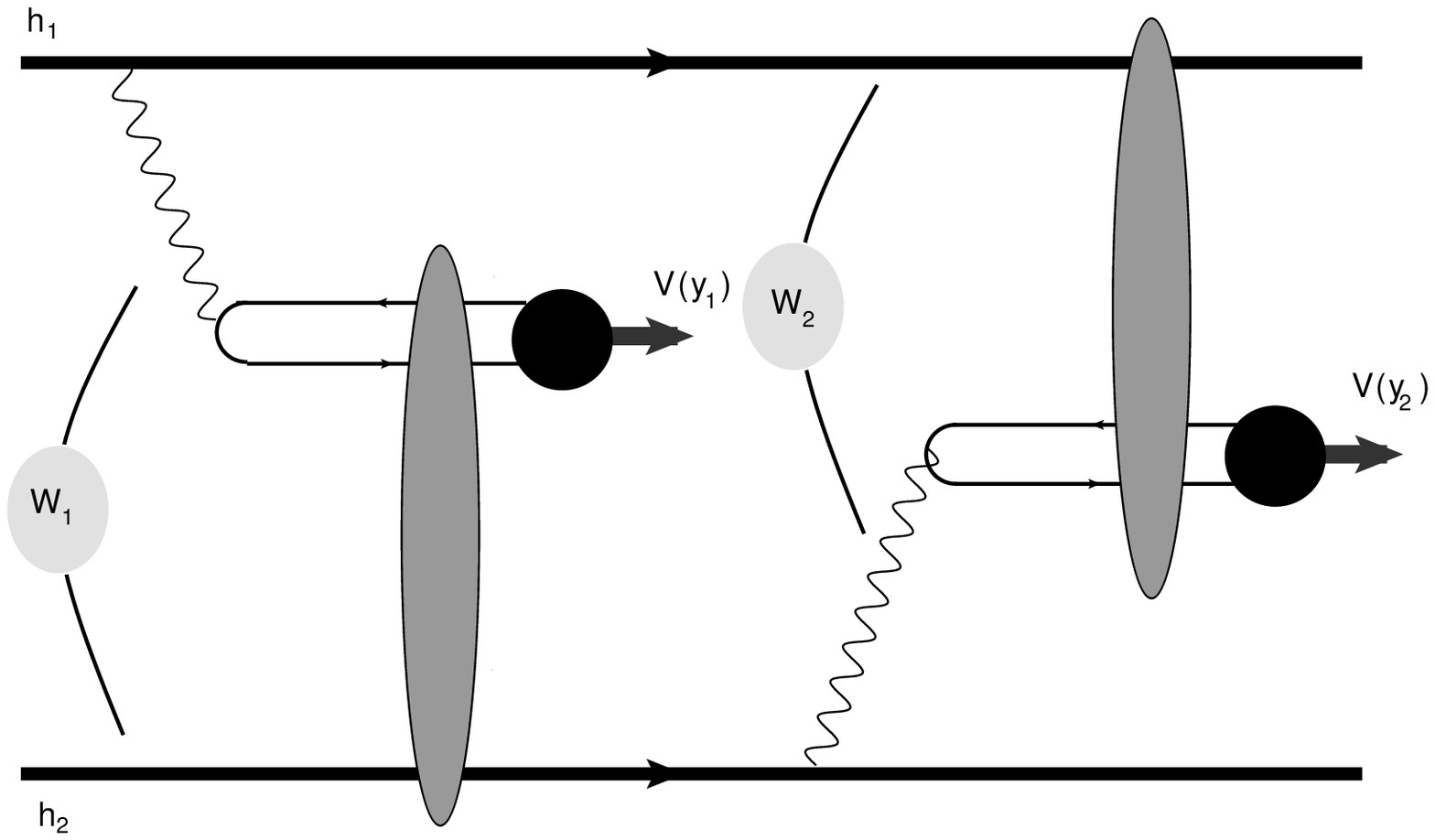,width=5.5cm}} & \,\,\,\,\,&  {\psfig{figure=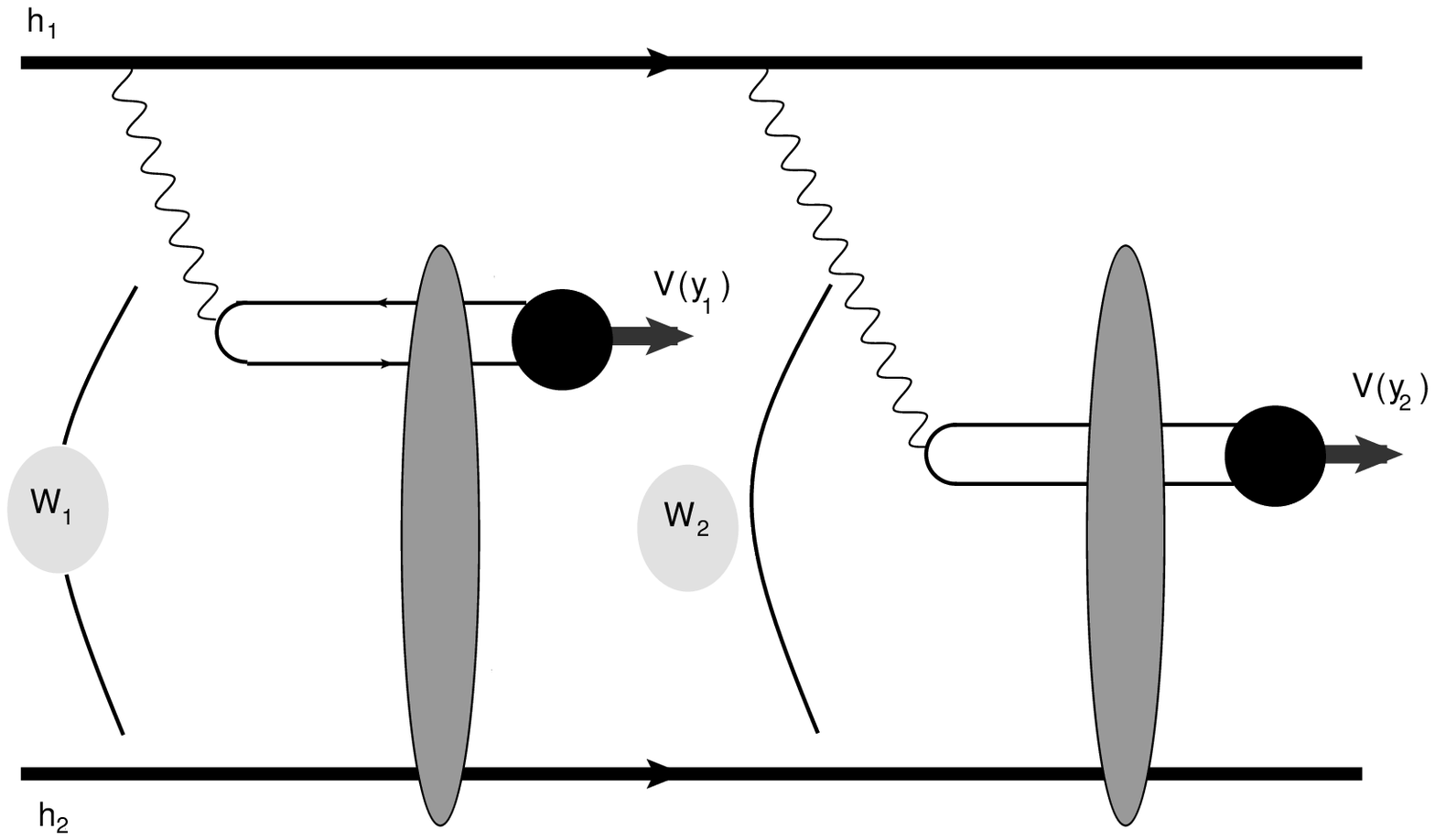,width=5.5cm}}  \\
\, & \, & \, \\
{\psfig{figure=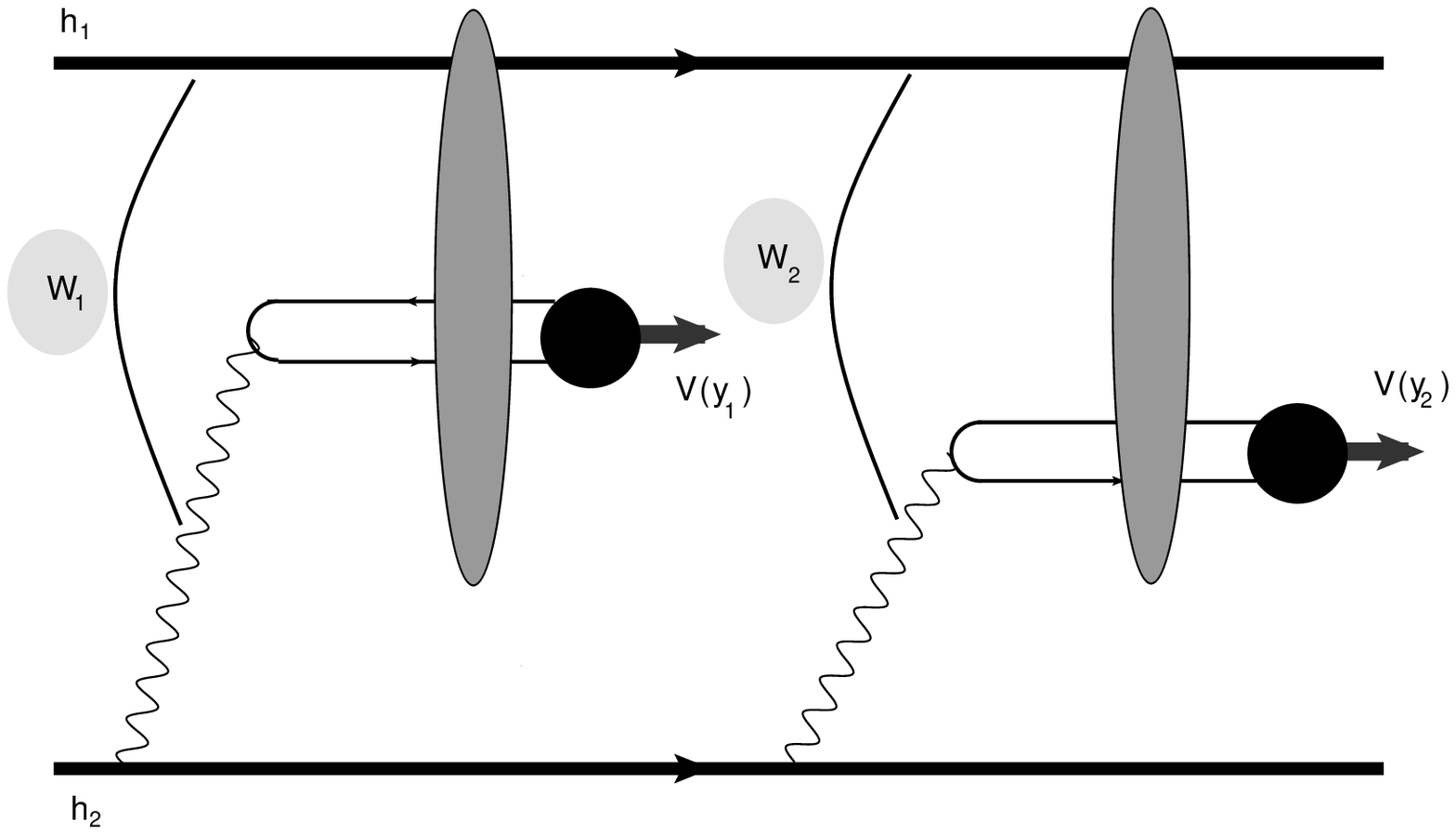,width=5.5cm}} & \,\,\,\,\,&{\psfig{figure=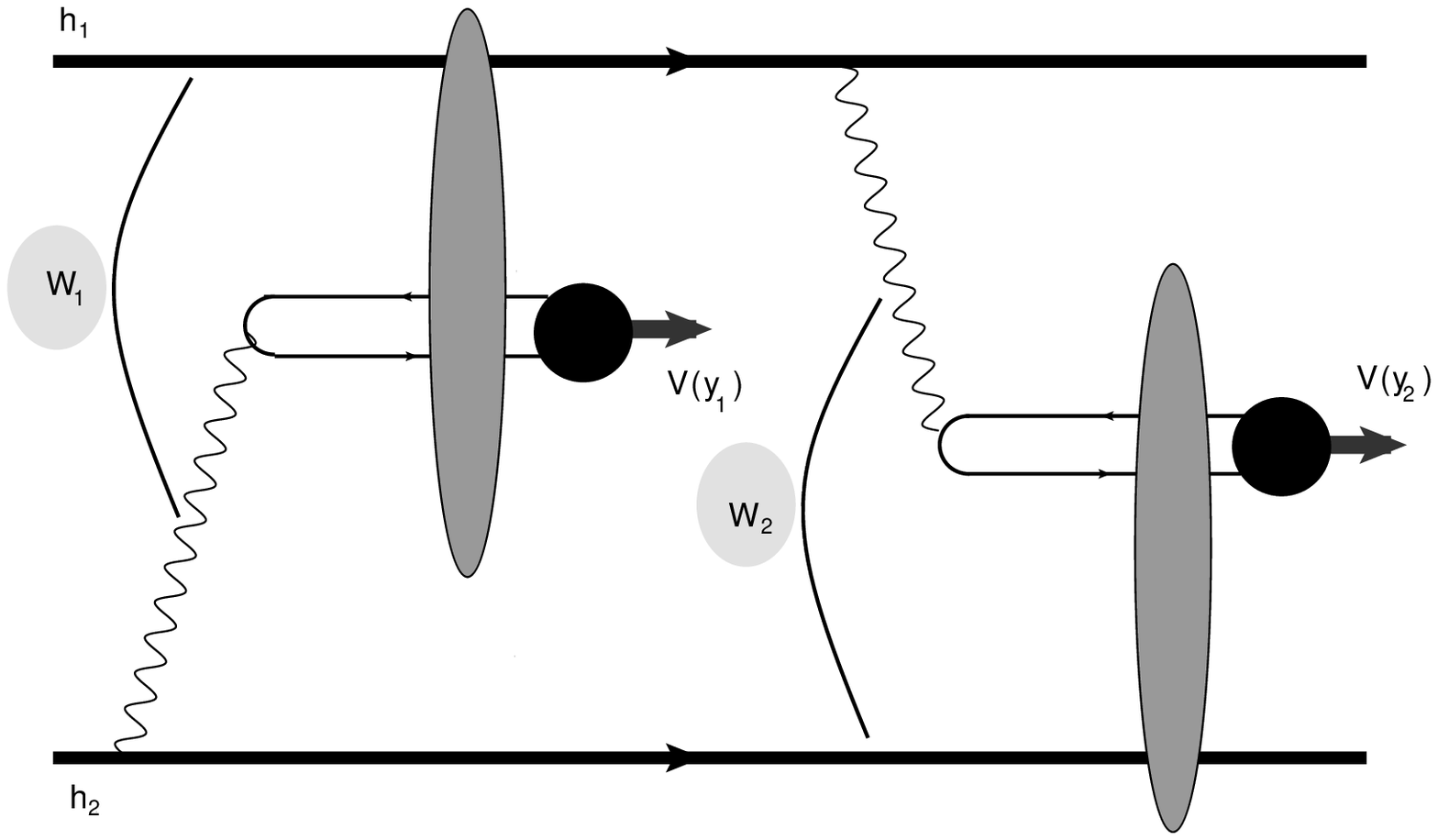,width=5.5cm}} 
\end{tabular}
%\centerline{\psfig{figure=gamavec.eps,width=10cm}}  
\caption{Double vector meson production in  $\gamma h$ interactions at hadronic colliders  in the 
color dipole picture.}
\label{dia2}
\end{figure}

Recent theoretical and experimental studies has demonstrated that hadronic colliders can also be considered photon -- hadron and photon -- photon colliders \cite{upc} which allow us to study the photon -- induced interactions in a new kinematical range and probe e.g. 
the nuclear gluon  distribution \cite{gluon,gluon2,gluon3,Guzey,vicwerluiz}, the dynamics of the strong interactions \cite{vicmag_mesons1,outros_vicmag_mesons,vicmag_update,motyka_watt,Lappi,griep,bruno1,bruno2}, the Odderon \cite{vicodd1,vicodd2}, the mechanism  of quarkonium production \cite{Schafer,mairon1,mairon2,cisek,bruno1,bruno2} and the photon flux of the proton \cite{vicgus1,vicgus2}. In particular, the installation of forward detectors  in the LHC \cite{ctpps,marek} should  allows to  separate more easily the exclusive processes, where the incident hadrons remain intact, allowing a detailed study of more complex final states as e.g. the exclusive production of two vector mesons explore other final states.  Recent results from the LHCb Collaboration for the exclusive double $J/\Psi$ production \cite{lhcb_dif} has demonstrate that the experimental analysis of this process is feasible, motivating the improvement of the theoretical description of this process \cite{kmr_duplo,vic_cris_dif,antonirhorho,antonipsipsi,bruno_doublegama}. In particular, in Ref. \cite{bruno_doublegama} we have revisited  the double vector production in $\gamma \gamma$ interactions, proposed originally in Refs. \cite{vicmagvv1,vicmagvv2,vicmagvv3}, taking into account recent improvements in the description of the $\gamma \gamma \rightarrow VV$ ($V = \rho, J/\Psi$) cross section  at low \cite{antonirhorho,antonipsipsi} and high \cite{brunodouble} energies. A typical diagram for this process is represented in Fig. \ref{dia1}. The results presented in Ref. \cite{bruno_doublegama}  has demonstrated that the analysis of this process is feasible in hadronic collisions, mainly in $pp$ collisions, and that its study  may be useful to constrain the QCD dynamics at high energies, as proposed originally in Ref. \cite{vicmagvv1}. However, double vector mesons  can also be produced in photon -- hadron ($\gamma h$) interactions if a double scattering occurs in a same event, as represented in Fig. \ref{dia2}. The treatment of this  double scattering mechanism (DSM) for $\gamma h$ interactions in heavy ion collisions was proposed originally in Ref. \cite{klein} and the double $\rho$ production was recently discussed in detail  in Ref. \cite{mariola}. Such results demonstrated that the contribution of the double scattering mechanism is important  for high energies, which motivates a more detailed analysis of this process.
In this paper we extend these previous studies for the double $J/\Psi$ and $\rho J/\Psi$ production in $AA$ collisions and present, by the first time, predictions for the double vector meson in $pp$ and $pA$ collisions. Additionally, we compare our results for double vector meson production in $\gamma h$ interactions with those obtained in Ref. \cite{bruno_doublegama} for $\gamma \gamma$ interactions. As we will demonstrate below, the $\rho\rho$ and  $J/\Psi J/\Psi$ production in $PbPb$ collisions is dominated by the double scattering mechanism, while the two - photon mechanism dominates in $pp$ collisions. Moreover, our results indicate that the analysis of the $\rho J/\Psi$ production at LHC can be useful to constrain the double scattering mechanism.

Lets start our analysis presenting a brief review of the main concepts and formulas to describe the single and double vector meson production in $\gamma h$ interactions at hadronic colliders.
The basic idea in the photon-induced processes is that
 a ultra relativistic charged hadron (proton or nucleus)
 give rise to strong electromagnetic fields, such that the photon stemming from the electromagnetic field of one of the two colliding hadrons can interact with one photon of
the other hadron (photon - photon process) or can interact directly with the other hadron (photon - hadron process) \cite{upc,epa}. In these processes the total cross section  can be factorized in
terms of the equivalent flux of photons into the hadron projectile and the photon-photon or photon-target production cross section. In this paper our main focus will be diffractive vector meson production in photon -- hadron interactions in hadronic collisions.   
The  differential cross sections for the production of a single vector meson $V$ at rapidity $y$ at fixed impact parameter $b$ of the hadronic collision can be expressed as follows:
\begin{eqnarray}
\frac{d\sigma \,\left[h_1 + h_2 \rightarrow   h_1 \otimes V \otimes h_2\right]}{d^2b dy} = \left[\omega N_{h_1}(\omega,b)\,\sigma_{\gamma h_2 \rightarrow V \otimes h_2}\left(\omega \right)\right]_{\omega_L} + \left[\omega  N_{h_2}(\omega,b)\,\sigma_{\gamma h_1 \rightarrow V \otimes h_1}\left(\omega \right)\right]_{\omega_R}\,
\label{dsigdy}
\end{eqnarray}
where the rapidity ($y$) of the vector meson in the final state is determined by the photon energy $\omega$ in the collider frame and by mass $M_{V}$ of the vector meson [$y\propto \ln \, ( \omega/M_{V})$]. Moreover, $\sigma_{\gamma h_i \rightarrow V \otimes h_i}$ is the total cross section for the diffractive vector meson photoproduction, with the symbol
$\otimes$ representing the presence of a rapidity gap in the final state and $\omega_L \, (\propto e^{-y})$ and $\omega_R \, (\propto e^{y})$ denoting  photons from the $h_1$ and $h_2$ hadrons, respectively.
One have that Eq. (\ref{dsigdy}) takes into account that both incident hadrons can be source of photon which will interact with the other hadron.  
The  equivalent photon 
spectrum $N(\omega,b)$ of a relativistic hadron for photons of energy $\omega$ at the distance ${\mathbf b}$ to the hadron trajectory, defined in the plane transverse to the trajectory, 
can be expressed in terms of the charge form factor $F$ as follows
\begin{eqnarray}
 N(\omega,b) = \frac{Z^{2}\alpha_{em}}{\pi^2}\frac{1}{b^{2}\omega}
\cdot \left[
\int u^{2} J_{1}(u) 
F\left(
 \sqrt{\frac{\left( \frac{b\omega}{\gamma_L}\right)^{2} + u^{2}}{b^{2}}}
 \right )
\frac{1}{\left(\frac{b\omega}{\gamma_L}\right)^{2} + u^{2}} \mbox{d}u
\right]^{2} \,\,,
\label{fluxo}
\end{eqnarray}
where $\gamma_L$ is the Lorentz factor. 
The double vector meson production can occur if two $\gamma h$ interactions   are present in the same event, as represented in Fig. \ref{dia2}. In order to treat this  double - scattering mechanism we will follow the approach from Refs. \cite{klein,mariola} that proposed that the double differential cross section for the production of a vector meson $V_1$ at rapidity $y_1$ and a second vector meson  $V_2$ at rapidity $y_2$ will be given by
\begin{eqnarray}
\frac{d^2\sigma_{h_1  h_2 \rightarrow   h_1 V_1 V_2 h_2}}{dy_1 dy_2} = {\cal{C}} \int_{b_{min}}
\frac{d\sigma \,\left[h_1 + h_2 \rightarrow   h_1  V_1  h_2\right]}{d^2b dy_1}
\times 
\frac{d\sigma \,\left[h_1 + h_2 \rightarrow   h_1  V_2  h_2\right]}{d^2b dy_2}
\,\, d^2b \,\,,
\label{Eq:double}
\end{eqnarray}  
where ${\cal{C}}$ is equal to 1 (1/2) for $V_1 \neq V_2$ ($V_1 = V_2$) and $b_{min} = R_{h_1} + R_{h_2}$ excludes the overlap between the colliding hadrons and allows to take into account only ultra peripheral collisions.  Consequently, the double vector meson production can be easily estimated in terms of the cross sections for the single vector meson production, which is determined by the photon flux and the $\gamma h \rightarrow V h$ cross section. 

In what  follows we will consider the color dipole formalism to describe the diffractive vector meson photoproduction, which successfully describe the HERA data and recent LHC data \cite{amir,bruno1,bruno2}. In this approach the description of the single vector meson production can be factorized as follows: i) a photon is emitted by one of the incident hadrons, ii) the photon fluctuates into  a quark-antiquark pair (the dipole), iii) this color dipole interact with the other hadron by the exchange of a color single state, denoted Pomeron ($I\!\!P$) and, iv) the pair converts into the vector meson final state.
The $\gamma h \rightarrow V h$  cross section is given by
\begin{eqnarray}
\sigma (\gamma h \rightarrow V h) =  \int_{-\infty}^0 \frac{d\sigma}{d{t}}\, d{t}  
= \frac{1}{16\pi}  \int_{-\infty}^0 |{\cal{A}}_T^{\gamma h \rightarrow V h }(x,\Delta)|^2 \, d{t}\,\,,
\label{sctotal_intt}
\end{eqnarray}
with the scattering amplitude is given by 
 \begin{eqnarray}
 {\cal A}_{T}^{\gamma h \rightarrow V h}({x},\Delta)  =  i
\int dz \, d^2\rr \, d^2\rb_h  e^{-i[\rb_h-(1-z)\rr].\rd} 
 \,\, (\Psi^{V*}\Psi)_{T}  \,\,2 {\cal{N}}_h({x},\rr,\rb_h) \,\,,
\label{sigmatot2}
\end{eqnarray}
where $(\Psi^{V*}\Psi)_{T}$ denotes the overlap of the transverse photon and vector meson wave functions. The variable  $z$ $(1-z)$ is the
longitudinal momentum fractions of the quark (antiquark) and  $\Delta$ denotes the transverse 
momentum lost by the outgoing pion (${t} = - \Delta^2$). 
The variable $\rb_h$ is the transverse distance from the center of the target $h$ to the center of mass of the $q \bar{q}$  dipole and the factor  in the 
exponential  arises when one takes into account 
non-forward corrections to the wave functions \cite{non}. As in our previous studies \cite{bruno1,bruno2} in what follows we will assume that  the vector meson is predominantly a quark-antiquark state 
and that the spin and polarization structure is the same as in the  photon \cite{dgkp,nnpz,sandapen,KT} (for other approaches see, for example, 
Ref. \cite{pacheco}). As a consequence, the overlap between the photon and the vector meson wave function, for the transversely  polarized  case, is given by (For details see Ref. \cite{KMW})
\begin{eqnarray}
  (\Psi^*_V\Psi)_T &=& \frac{\hat e_fe}{4\pi}\frac{N_c}{\pi z(1-z)}
    \left\{m_f^2K_0(\epsilon r)\phi_T(r,z)-\left[z^2+(1-z)^2\right]\epsilon K_1(\epsilon r)\partial_r\phi_T(r,z)\right\}, 
\end{eqnarray}
where $ \hat{e}_f $ is the effective charge of the vector meson, $m_f$ is the quark mass, $N_c = 3$, $\epsilon^2 = z(1-z)Q^2 + m_f^2$   
and $\phi_i(r,z)$ define the scalar parts of the  vector meson wave functions. In the   
Gauss-LC model one have that
\begin{eqnarray}
  \phi_T(r,z) &=& N_T\left[z(1-z)\right]^2\exp\left(-r^2/2R_T^2\right)\,.
\end{eqnarray}
with the parameters  $N_T$  and $R_T$ being   determined by the normalization condition of the wave function and by the meson decay width (For details see Table 1 in Ref. \cite{bruno2}). It is important to emphasize that predictions 
based on this model for the wave functions have been tested with success in $ep$ and ultra peripheral hadronic collisions (See, e. g. 
Refs. \cite{amir,anelise,bruno1,bruno2}).
 Moreover, ${\cal{N}}_h (x,\rr,\rb_h)$ denotes the non-forward scattering  amplitude of a dipole of size $\rr$ on the hadron $h$, which is  directly related to  the QCD 
dynamics. In what follows we will assume that for the proton case  
${\cal{N}}_p (x,\rr,\rb_p)$ is given by the 
bCGC model proposed in Ref. \cite{KMW}, which improves the Iancu - Itakura - Munier (IIM) model 
 \cite{iim} with  the inclusion of   the impact parameter dependence in the dipole - proton scattering amplitude.   Following \cite{KMW} we have:
\begin{eqnarray}
\mathcal{N}_p(x,\rr,{\rb_p}) =   
\left\{ \begin{array}{ll} 
{\mathcal N}_0\, \left(\frac{ r \, Q_{s,p}}{2}\right)^{2\left(\gamma_s + 
\frac{\ln (2/r Q_{s,p})}{\kappa \,\lambda \,Y}\right)}  & \mbox{$r Q_{s,p} \le 2$} \\
 1 - \exp \left[-A\,\ln^2\,(B \, r \, Q_{s,p})\right]   & \mbox{$r Q_{s,p}  > 2$} 
\end{array} \right.
\label{eq:bcgc}
\end{eqnarray}
with  $Y=\ln(1/x)$ and $\kappa = \chi''(\gamma_s)/\chi'(\gamma_s)$, where $\chi$ is the 
LO BFKL characteristic function \cite{bfkl}.  The coefficients $A$ and $B$  
are determined uniquely from the condition that $\mathcal{N}_p(x,\rr,\rb_p)$, and its derivative 
with respect to $rQ_s$, are continuous at $rQ_s=2$. 
In this model, the proton saturation scale $Q_{s,p}$ depends on the impact parameter:
\begin{equation} 
  Q_{s,p}\equiv Q_{s,p}(x,{\rb_p})=\left(\frac{x_0}{x}\right)^{\frac{\lambda}{2}}\;
\left[\exp\left(-\frac{{b_p}^2}{2B_{\rm CGC}}\right)\right]^{\frac{1}{2\gamma_s}}.
\label{newqs}
\end{equation}
The parameter $B_{\rm CGC}$  was  adjusted to give a good 
description of the $t$-dependence of exclusive $J/\psi$ photoproduction.  
The factors $\mathcal{N}_0$, $x_0$,  $\lambda$ and  $\gamma_s$  were  taken  to be free. Recently the parameters of this model have been updated in Ref. \cite{amir} (considering the 
recently released high precision combined HERA data), giving  $\gamma_s = 0.6599$, $B_{CGC} = 5.5$ GeV$^{-2}$,
$\mathcal{N}_0 = 0.3358$, $x_0 = 0.00105 \times 10^{-5}$ and $\lambda = 0.2063$. As demonstrate in Ref. \cite{armesto_amir}, this phenomenological dipole  describes quite well the HERA data for the exclusive $\rho$ and $J/\Psi$ production. Moreover, the results from Refs. \cite{bruno1,bruno2} 
demonstrated that this model allows to describe the recent LHC data for the exclusive vector meson photoproduction in $pp$ and $pPb$ collisions. Another motivation to use the bCGC model, is that this model is based on the CGC physics, which  was  used in Ref. \cite{bruno_doublegama} to estimate the double vector meson production in $\gamma \gamma$ interactions. A common approach for the QCD dynamics in $\gamma \gamma$ and $\gamma h$ interactions is important to minimize the theoretical uncertainty and to perform a realistic comparison between the  predictions of the two different mechanisms for the double vector production.
In order to describe the vector meson production in $\gamma A$ interactions we need to specify the 
forward dipole - nucleus scattering amplitude, $\mathcal{N}_A(x,\rr,\rb_A)$.  
Following \cite{bruno1} we will use in our calculations  the model proposed in Ref. 
\cite{armesto}, which describes  the current  experimental data on the nuclear 
structure function as well as includes the  impact parameter dependence in the dipole 
nucleus cross section. In this model the forward dipole-nucleus amplitude is given by
\begin{eqnarray}
{\cal{N}}_A(x,\rr,\rb_A) = 1 - \exp \left[-\frac{1}{2}  \, \sigma_{dp}(x,\rr^2) 
\,T_A(\rb_A)\right] \,\,,
\label{enenuc}
\end{eqnarray}
where $\sigma_{dp}$ is the dipole-proton cross section given by
\begin{eqnarray}
\sigma_{dp} (x,\rr^2) = 2 \int d^2\rb_p \,\,\mathcal{N}_p(x,\rr,{\rb_p})  
\end{eqnarray}
 and $T_A(\rb_A)$ is the nuclear profile 
function, which is obtained from a 3-parameter Fermi distribution for the nuclear
density normalized to $A$.
The above equation
%, based on the Glauber-Gribov formalism \cite{gribov},  
sums up all the 
multiple elastic rescattering diagrams of the $q \overline{q}$ pair
and is justified for large coherence length, where the transverse separation $\rr$ of 
partons 
in the multiparton Fock state of the photon becomes a conserved quantity, {\it i.e.} 
the size 
of the pair $\rr$ becomes eigenvalue
of the scattering matrix.

In the case of the double vector meson production in $\gamma \gamma$ interactions at hadronic colliders, represented in Fig. \ref{dia1}, we have that the total cross section is given by (For details see Ref. \cite{bruno_doublegama})
\begin{eqnarray}
\sigma \left( h_1 h_2 \rightarrow h_1 \otimes V_1V_2 \otimes h_2 ;s \right)   
&=& \int \hat{\sigma}\left(\gamma \gamma \rightarrow V_1V_2 ; 
W_{\gamma \gamma} \right )  N\left(\omega_{1},{\mathbf b_{1}}  \right )
 N\left(\omega_{2},{\mathbf b_{2}}  \right ) S^2_{abs}({\mathbf b})  
\frac{W_{\gamma \gamma}}{2} \mbox{d}^{2} {\mathbf b_{1}}
\mbox{d}^{2} {\mathbf b_{2}} 
\mbox{d}W_{\gamma \gamma} 
\mbox{d}Y \,\,\, .
\label{cross-sec-2}
\end{eqnarray}
where  $\omega_1$ and $\omega_2$ are the photon energies, 
$W_{\gamma \gamma} = \sqrt{4 \omega_1 \omega_2}$ is the invariant mass of the $\gamma \gamma$ system and   $Y$ is the rapidity of the outgoing double meson system. Moreover, $S^2_{abs}({\mathbf b})$ is the absorption factor, given in what follows by
\begin{eqnarray}
S^2_{abs}({\mathbf b}) = \Theta\left(
\left|{\mathbf b}\right| - R_{h_1} - R_{h_2}
 \right )  = 
\Theta\left(
\left|{\mathbf b_{1}} - {\mathbf b_{2}}  \right| - R_{h_1} - R_{h_2}
 \right )  \,\,,
\label{abs}
\end{eqnarray}
where $R_{h_i}$ is the radius of the hadron $h_i$ ($i = 1,2$). In the dipole picture, the $\gamma \gamma \rightarrow V_1 V_2$ cross section can be expressed as follows
\begin{eqnarray}
\sigma\, (\gamma \gamma \rightarrow V_1 \, V_2)  =  \frac{[{\cal I}m \, {\cal A}(W_{\gamma \gamma}^2,\,t=0)]^2}{16\pi\,B_{V_1 \,V_2}} \;,
\label{totalcs}
\end{eqnarray}
where we have  approximated the $t$-dependence of the differential cross section by an exponential  with  $B_{V_1 \, V_2}$ being 
the slope parameter. The imaginary part of the amplitude at zero momentum transfer ${\cal A}(W_{\gamma \gamma}^2,\,t=0)$ reads as
\begin{eqnarray}
{\cal I}m \, {\cal A}\, (\gamma \gamma \rightarrow V_1 \, V_2) & = &  
\int dz_1\, d^2\rr_1 \,[\Psi^\gamma(z_1,\,\rr_1)\,\, \Psi^{V_1*}(z_1,\,\rr_1)]_T \nonumber \\
&\times & \int dz_2\, d^2\rr_2 \,[\Psi^\gamma(z_2,\,\rr_2)\,\, \Psi^{V_2 *}(z_2,\,\rr_2)]_T
\,
\sigma_{d d}(\rr_1, \rr_2,Y)
 \, ,
\label{sigmatot}
\end{eqnarray}
where $\Psi^{\gamma}$ and $\Psi^{V_i}$  are the light-cone wave functions  of the photon and vector meson, respectively, and $T$ the transverse polarization.  The variable $\rr_1$ defines the relative transverse
separation of the pair (dipole) and $z_1$ $(1-z_1)$ is the longitudinal momentum fraction of the quark (antiquark). Similar definitions are valid 
for $\rr_2$ and  $z_2$. Moreover, $\sigma_{d d}$ is the  dipole - dipole cross section, which can be estimated taking into account the nonlinear effects in the QCD dynamics. In what follows, we assume the Gauss-LC model for the vector meson wave functions and   estimate $\sigma_{d d}$  using the approach proposed  in Refs. \cite{nosfofo,bruno_doublegama}, which is based on the CGC physics. We refer the reader to the Ref. \cite{bruno_doublegama} for more details about the double vector meson production in $\gamma \gamma$ interactions.

\begin{figure}[t]
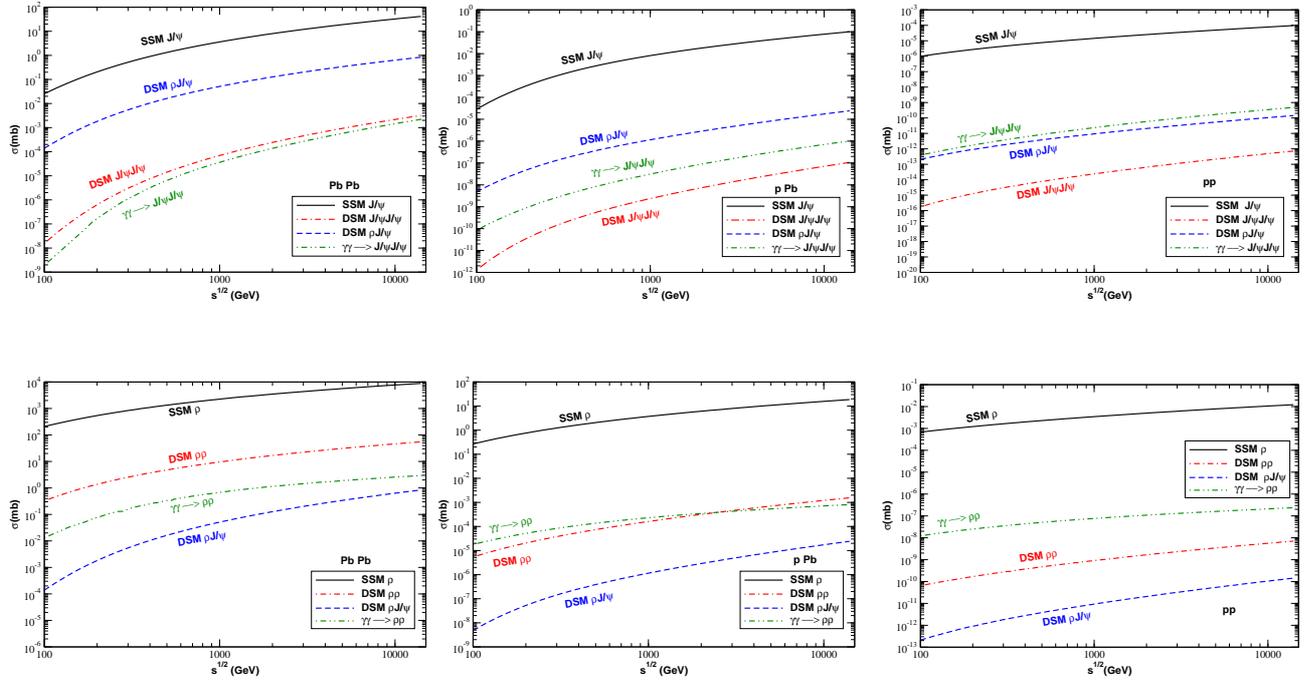

\begin{tabular}{cccc}
{\psfig{figure=new_AA-loopenergy-psi-r.eps,width=5.5cm}} & {\psfig{figure=new_pA-loopenergy-psi-r.eps,width=5.5cm}} & \, & {\psfig{figure=new_pp-loopenergy-psi-r.eps,width=5.5cm}} \\
\vspace{0.5cm}\, & \, & \, \\
{\psfig{figure=new_AA-loopenergy-rho-r.eps,width=5.5cm}} &{\psfig{figure=new_pA-loopenergy-rho-r.eps,width=5.5cm}}  &\, &{\psfig{figure=new_pp-loopenergy-rho-r.eps,width=5.5cm}} 
\end{tabular}
%\centerline{\psfig{figure=gamavec.eps,width=10cm}}  
%\centerline{\psfig{figure=diagrama.eps,width=10cm}}
\caption{Energy dependence for the  $J/\Psi J/\Psi$ (upper panels) and  $\rho \rho$ (lower panels) production in  $\gamma h$ and $\gamma \gamma$  interactions in $PbPb$ (left panels), $pPb$ (middle panels) and $pp$ (right panels) collisions. The predictions associated to the double scattering mechanism are denoted DSM. For comparison the predictions for the single vector meson production, denoted SSM, are also shown. }
\label{fig2}
\end{figure}

\begin{table}[t] % aqui começa o ambiente tabela
\centering
\begin{tabular}{||c|c|c|c|c|c|c||} 
\hline 
\hline
Final state & Mechanism & $PbPb$ & $PbPb$ & $pPb$ & $pp$ & $pp$ \\
\, & \, & $\sqrt{s}=2.76\,\mbox{TeV}$ & $\sqrt{s}=5.5\,\mbox{TeV}$ & $\sqrt{s}=5\,\mbox{TeV}$ & $\sqrt{s}=7\,\mbox{TeV}$ & $\sqrt{s}=14\,\mbox{TeV}$ \\
\hline
\hline
$J/\Psi J/\Psi$ & DSM & 402.301 nb & 1054.951 nb & 28.473 pb & 3.223 $\times$10$^{-4}$ pb & 7.256$\times$10$^{-4}$ pb \\
\, & $\gamma \gamma$ &  235.565 nb & 658.589 nb  &  310.194 pb  & 0.2412 pb  &  0.4793 pb \\
\hline
\hline
$\rho \rho$ & DSM & 21.150 mb & 29.421 mb & 702.595 nb & 4.354 pb & 7.083 pb \\
\, & $\gamma \gamma$ &  1.389 mb & 1.973 mb &  536.432 nb  & 182.442 pb  & 237.006 pb \\
\hline
\hline
$\rho J/\Psi$ & DSM & 0.18 mb & 0.35 mb & 8.929 nb & 7.469 $\times$10$^{-2}$ pb & 14.288 $\times$10$^{-2}$ pb \\
\hline
\hline
\end{tabular}
\caption{Total cross sections for the double vector meson production considering the double scattering and  two - photon mechanisms and different center - of - mass energies considering the full kinematical range covered by the LHC.} % igual ao ambiente figura
\label{tab1}
\end{table}

In what follows we present our predictions for the rapidity distributions and total cross sections for the $\rho \rho$, $\rho J/\Psi$ and $J/\Psi J/\Psi$ production in $\gamma h$ interactions at $pp/pPb/PbPb$ collisions. We will denote the predictions associated to the double scattering mechanics by DSM hereafter. Following Ref. \cite{mariola} we will estimate the  equivalent photon spectra for $A = Pb$ assuming the nucleus as a point - like object, i.e. $F(q^2) = 1$. In the proton case, we will take   $F(q^2) = 1/[1 + q^2/(0.71 \mbox{GeV}^2)]^2$ and $R_p = 0.7$ fm as in Ref. \cite{bruno_doublegama}. Moreover, we will compare our predictions for the  $J/\Psi J/\Psi$ and $\rho \rho$ production with the results obtained in Ref. \cite{bruno_doublegama} for the production of these final states in $\gamma \gamma$ interactions. In Fig. \ref{fig2} we present our predictions for the energy dependence of the total cross sections for the double vector meson production in $\gamma h$ and $\gamma \gamma$  interactions. For the double $J/\Psi$ production (upper panels), the double scattering mechanism becomes competitive with the two - photon one only in $PbPb$ collisions, being a factor 10 (100) smaller in $pPb$ ($pp$) collisions. In particular, for $pp$ collisions, the DSM contribution is negligible.  On the other hand, our results demonstrate that the associated production of a $J/\Psi$ and a $\rho$ meson by the double scattering mechanism is important the LHC range. It is important to emphasize that  this final state also can be produced by $\gamma \gamma$ interactions. However, as its contribution in hadronic collisions still is an open question due to the current large uncertainty on the normalization of the $\gamma \gamma \rightarrow \rho J\Psi$ cross section (For a detailed discussion see Ref. \cite{brunodouble}), we do not present the associated predictions.  In the case of the double $\rho$ production (lower panels), the double scattering mechanism is dominant in $PbPb$ collisions, in agreement with the results presented in Ref. \cite{mariola}. On the other hand, the contribution of the double scattering and two - photon mechanisms are similar in $pPb$ collisions, while the $\gamma \gamma$ dominates in the $pp$ collisions.   These results demonstrate that the analysis of this final state in $PbPb / pPb / pp$ can be useful to disentangle the different mechanisms for the $\rho \rho$ production. The corresponding total cross sections at different values of the center - of - mass energy are presented in Table \ref{tab1} considering the full kinematical range covered by the LHC.

\begin{figure}
\begin{tabular}{ccc}
{\psfig{figure=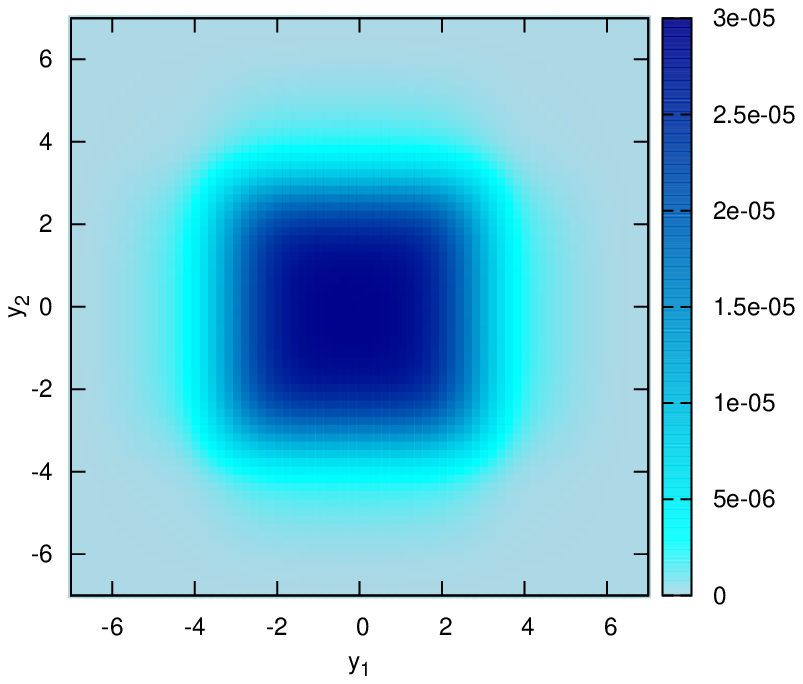,width=6cm}} &
{\psfig{figure=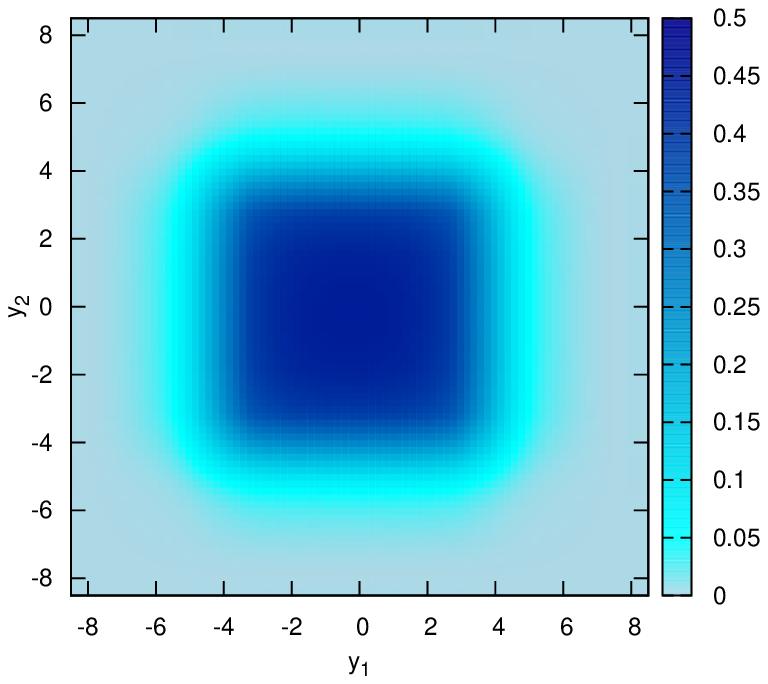,width=6cm}} &
{\psfig{figure=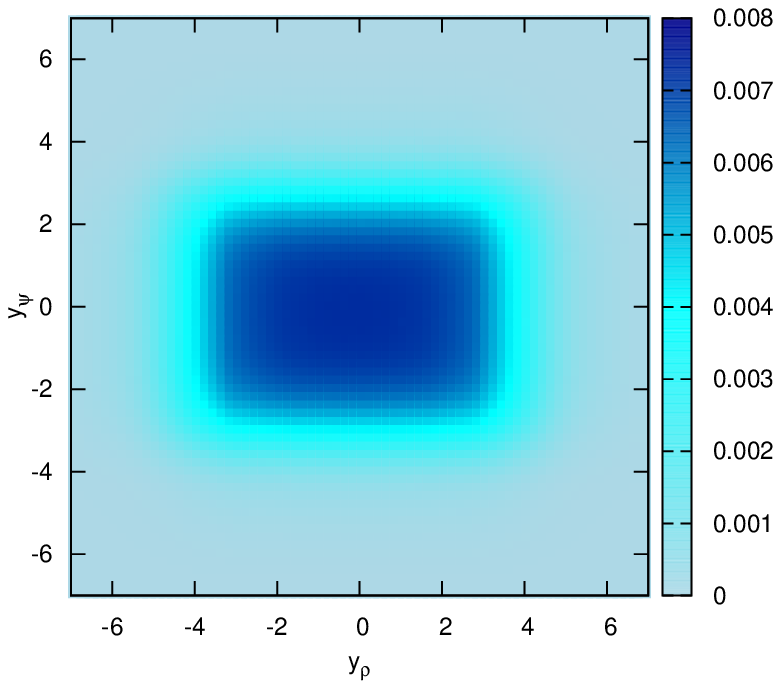,width=6cm}}
\end{tabular}                                                                                                                       
\caption{Double differential rapidity distribution for the $J/\Psi J/\Psi$ (left panel), $\rho \rho$ (middle panel) and $\rho J/\Psi$ (right panel)  production in $\gamma h$ interactions at $PbPb$ collisions ($\sqrt{s} = 5.5$ TeV) by the double scattering mechanism.}
\label{fig3}
\end{figure}

\begin{figure}
\begin{tabular}{ccc} 
{\psfig{figure=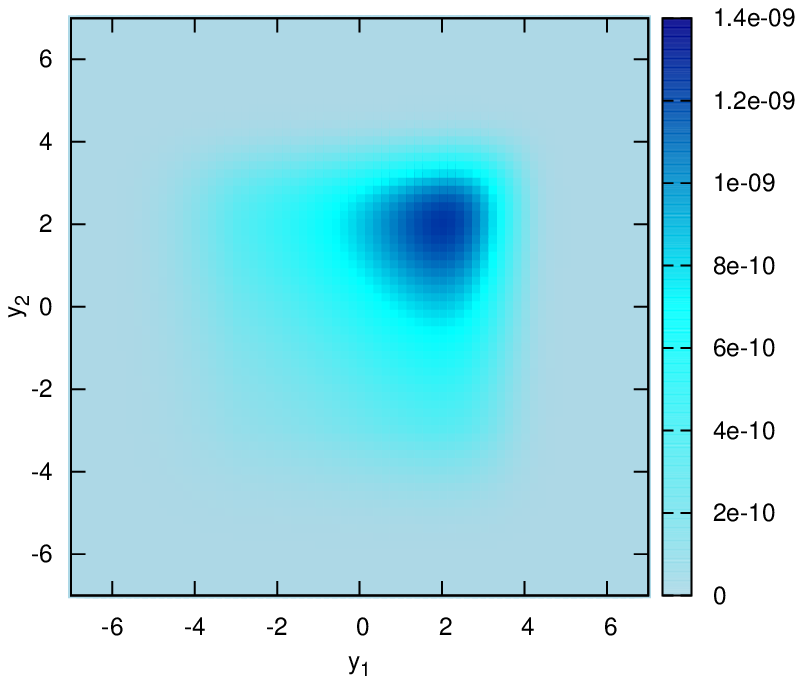,width=6cm}} &
{\psfig{figure=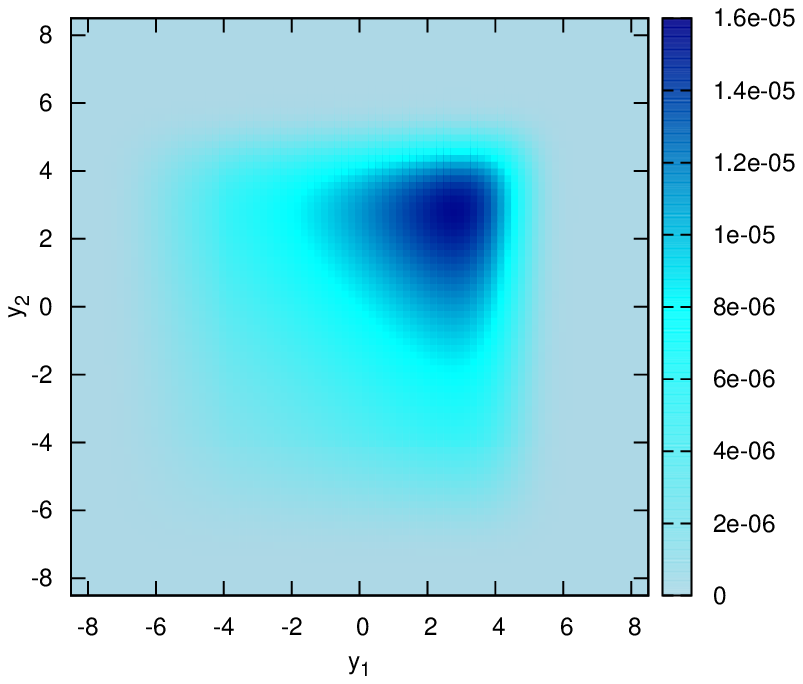,width=6cm}} &
{\psfig{figure=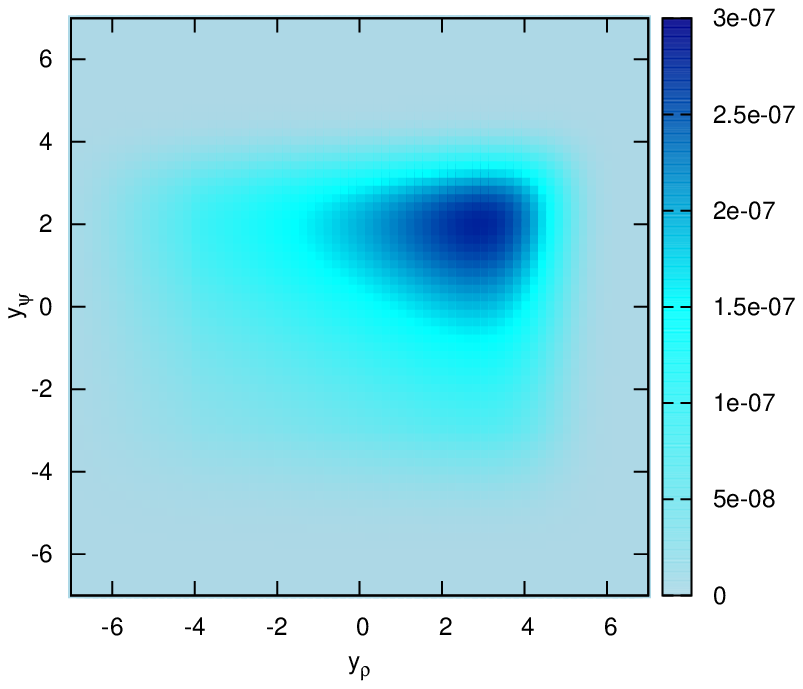,width=6cm}}
\end{tabular}                                                                                                                       
\caption{Double differential rapidity distribution for the $J/\Psi J/\Psi$ (left panel), $\rho \rho$ (middle panel) and $\rho J/\Psi$ (right panel) production in $\gamma h$ interactions at $pPb$ collisions ($\sqrt{s} = 5$ TeV) by the double scattering mechanism.}
\label{fig4}
\end{figure}

In Figs. \ref{fig3} and \ref{fig4} we present our predictions for the rapidity distributions for the double vector meson production by the double scattering mechanism in $PbPb$ and $pPb$ collisions, respectively. For $PbPb$ collisions, as expected, one have symmetric distributions for the  $J/\Psi J/\Psi$ and $\rho \rho$ production. On the other hand, in the case of the  $\rho J/\Psi$ production, the distribution is asymmetric, being wider for the rapidity associated to the $\rho$ meson. In the case of $pPb$ collisions, one have that the photon flux of the nucleus is amplified by a factor $Z^2$ in comparison to the photon flux associated to the proton. As a consequence, the  double scattering mechanism is dominated by $\gamma h$ interactions where the photons are emitted by the nucleus. The contribution associated to one photon emitted by the nucleus and the other by the proton is suppressed by a factor $Z^2$, while the contribution associated to $\gamma h$ interactions with  photons emitted by the proton is suppressed by a factor $Z^4$. It implies that the rapidity distributions are asymmetric for all final states considered (See Fig. \ref{fig4}). Similarly as observed in $PbPb$ collisions, the rapidity distribution associated to the $\rho$ meson is wider in comparison to the $J/\Psi$ one.

Finally, in Table \ref{tab2} we present our predictions for the total cross sections for the double vector production by the double scattering mechanism in $PbPb$ and $pPb$ collisions considering the rapidity ranges covered by the ATLAS, CMS, ALICE and LHCb Collaborations. In the particular case of the ALICE Collaboration we estimate the cross sections considering: (a) that both mesons are produced in the range $-1 < y_{1,2} < 1$ (denoted ALICE1 in the Table) and (b) that one meson is produced in the range 
$-1 < y_{1} < 1$ and the other in the range $-3.6 < y_{2} < -2.6$ (denoted ALICE2). For the $\rho J/\Psi$ production in the ALICE2 range, we present our results for the two possible configurations: $(y_1, y_2) = (y_{\rho}, y_{J/\Psi})$ and $(y_1, y_2) = (y_{J/\Psi},y_{\rho})$, with the results associated to the latter one being presented in parenthesis in Table \ref{tab2}. We predict large values for the total cross sections, in particular, for the $\rho \rho$ and $\rho J/\Psi$ production in $PbPb$ collisions,  in the phase space covered by the different collaborations. Consequently, we believe that the analysis of these different final states is feasible in the future, which will allow to probe the double scattering mechanism at the LHC.

\begin{table}[t] % aqui começa o ambiente tabela
\centering
\begin{tabular}{||c|c|c|c|c|c||} 
\hline 
\hline
Final state & \, &   LHCb & ATLAS/CMS & ALICE1 & ALICE2 \\
\, & \, & $2 < y_{1,2} < 4.5$ & $-2 < y_{1,2} < 2$ & $-1 < y_{1,2} < 1$ &  $-1 < y_{1} < 1$ and $-3.6 < y_{2} < -2.6$ \\
\hline
\hline
$J/\Psi J/\Psi$ &  $PbPb \,\, (\sqrt{s}=2.76\,\mbox{TeV})$ & 5.51 nb  & 234.94 nb   & 69.91 nb   & 6.94 nb  \\  
 \, &   $PbPb \,\, (\sqrt{s}=5.5\,\mbox{TeV})$  & 30.85 nb & 446.11 nb   & 118.03 nb  & 25.45 nb \\     \, & $pPb \,\, (\sqrt{s}=5\,\mbox{TeV})$     & 3.25 pb  & 8.87 pb     & 2.16 pb    & 0.37 pb  \\ 
\hline
\hline
$\rho \rho$ &   $PbPb \,\, (\sqrt{s}=2.76\,\mbox{TeV})$ &  0.93 mb  & 6.08 mb   & 1.58 mb & 0.54 mb \\  
  \, &   $PbPb \,\, (\sqrt{s}=5.5\,\mbox{TeV})$  &  1.50 mb  & 7.06 mb   & 1.79 mb & 0.73 mb  \\ 
    \, &   $pPb \,\, (\sqrt{s}=5\,\mbox{TeV})$     & 84.09 nb  & 122.03 nb & 30.11 nb& 8.53 nb   \\  \hline
\hline
$\rho J/\Psi$ & $PbPb \,\, (\sqrt{s}=2.76\,\mbox{TeV})$ &  4.48 $\mu$b  & 75.17 $\mu$b   & 20.94 $\mu$b   & 2.06 (7.25) $\mu$b  \\  
  \, &    $PbPb \,\, (\sqrt{s}=5.5\,\mbox{TeV})$  &  13.42 $\mu$b & 112.00 $\mu$b  & 29.06 $\mu$b   & 6.21 (11.86) $\mu$b \\ 
  \, &    $pPb \,\, (\sqrt{s}=5\,\mbox{TeV})$     &  1.02 nb      &  2.08 nb       &  0.51 nb       & 87.31 (144.56) pb \\
\hline
\hline
\end{tabular}
\caption{Total cross sections for the double vector meson production by the double scattering mechanism (DSM) for different center - of - mass energies considering the distinct phase space in rapidity covered by the  ALICE, ATLAS, CMS and LHCb Collaborations.} % igual ao ambiente figura
\label{tab2}
\end{table}

Let us summarize our main conclusions. In recent years, a series of studies have discussed in detail the treatment of the total cross section and 
the exclusive production of different final states in $\gamma \gamma$ and $\gamma h$ interactions considering very distinct theoretical approaches.  
In particular, recent results for the double vector meson production in $\gamma \gamma$ interactions at hadronic colliders has demonstrated that this process can be used to constrain the 
QCD dynamics at high energies. However, this final state can  also be generated if double $\gamma h$ interactions are present in the same event. In this paper we have estimated the magnitude of this contribution for the $J/\Psi J/\Psi$, $\rho \rho$ and $\rho J/\Psi$ production in $PbPb/pPb/pp$ collisions. We have treated the double scattering and two - photon mechanisms using  the dipole formalism and a same approach for  the QCD dynamics and the vector meson wave function. Our results indicated that the DSM contribution  is dominant for the $J/\Psi J/\Psi$ and $\rho \rho$ production in $PbPb$ collisions. On the other hand, the two - photon production dominates the double $J/\Psi$ production in $pPb$ and $pp$ collisions. In the case of the double $\rho$ production, the DSM and two - photon contributions are similar in $pPb$ collisions, with the two - photon being dominant in $pp$ collisions. Therefore, the analysis of double vector production considering different projectiles can be useful to disentangle the different mechanisms of production. In particular, the analysis of the DPS production in heavy ion collisions can be used to complement our understanding of the description of the diffractive  vector meson photoproduction. Moreover, our results demonstrated that the DPS $\rho J/\Psi$ production is large in the LHC kinematical range. Finally, our predictions for the double vector meson production in the phase space covered by the different experimental collaborations at the LHC indicate that the study of the double vector meson production is feasible in the future.

{\it Note added in the proof:}  One month after the submission of this paper, a report has appeared \cite{LNS} where it has been estimated  the exclusive double $\rho$ production in $pp$ collisions. The total cross section was estimated in \cite{LNS} taking into account pomeron and reggeon exchanges and considering the tensor pomeron model proposed in Ref. \cite{N1} and discussed in detail in Ref. \cite{N2}.  The cross sections 
found in \cite{LNS} are more than three orders of magnitude larger than our predictions. Therefore, the double $\rho$ production in $pp$ collisions is predicted to be dominated by pomeron - pomeron interactions, which implies that the analysis of this process can be useful to probe the tensor pomeron model. An alternative to study the photon - induced $\rho \rho$ production in $pp$ collisions analysed here  is the reconstruction of the entire event with a cut on the summed transverse momentum
of the event. As the typical photon virtualities
are very small, the hadron scattering angles are very
low. Consequently, we expect a different transverse
momentum distribution of the scattered hadrons, with pomeron - pomeron 
interactions predicting larger $p_T$ values. Surely this subject deserve a more detailed analysis in the future. Finally, it is important to emphasize that  in contrast to $pp$ collisions, the photon - induced interactions are expected to be dominant in $pA (AA)$ collisions due to the $Z^2 \, (Z^4)$   enhancement associated to the presence of nuclear photon flux.

\begin{acknowledgments}
VPG thanks G. Contreras, S. Klein,  R. McNulty and D. Tapia Takaki by useful discussions. 
This work was  partially financed by the Brazilian funding agencies CNPq, CAPES, FAPERGS and FAPESP.

\end{acknowledgments}

\hspace{1.0cm}

\end{document}